%% file: main.tex
\documentclass[sigconf,10pt,nonacm]{acmart}
 
\AtBeginDocument{%
  \providecommand\BibTeX{{%
    \normalfont B\kern-0.5em{\scshape i\kern-0.25em b}\kern-0.8em\TeX}}}

\usepackage[utf8]{inputenc}
\usepackage{color}
\usepackage{textcomp}
\usepackage{url}
\usepackage{amsmath}
\usepackage{graphicx}
\usepackage{balance}
\usepackage{booktabs}
\usepackage[font=small]{caption}

\begin{document}
\title{Improving Aircraft Localization: Experiences and Lessons Learned from an Open Competition}

\author{Martin Strohmeier}
\email{martin.strohmeier@armasuisse.ch}
\orcid{0000-0002-1936-0933}
\affiliation{%
 \institution{Cyber-Defence Campus, armasuisse Science + Technology}
 \city{Zurich}
 \country{Switzerland}
}

\author{Mauro Leonardi}
\email{mauro.leonardi@uniroma2.it}

\affiliation{%
 \institution{University of Rome Tor Vergata}
 \city{Rome}
 \country{Italy}}

\author{Sergei Markochev}
\email{sergey.markochev@gmail.com}

\affiliation{%
 \institution{Independent}
 \city{London}
 \country{United Kingdom}
}

\author{Fabio Ricciato}
\email{fabio.ricciato@gmail.com}
\affiliation{%
 \institution{Independent}
  \country{Luxembourg}}

\author{Matthias Schäfer}
\email{schaefer@sero-systems.de}

\affiliation{%
 \institution{Sero Systems}
 \city{Kaiserslautern}
 \country{Germany}}

\author{Vincent Lenders}
\email{vincent.lenders@armasuisse.ch}

\affiliation{%
 \institution{Cyber-Defence Campus, armasuisse Science + Technology}
 \city{Zurich}
 \country{Switzerland}
}

\renewcommand{\shortauthors}{Strohmeier et al.}

\begin{abstract}
Knowledge about the exact positioning of aircraft is crucial in many settings, both in times of war and peace. Consequently, the opportunistic and independent localization of aircraft based on their communication has been a longstanding problem and subject of much research. Originating from military settings, the capability to conduct aircraft localization has moved first towards the institutional civil aviation domain and can now be undertaken by anyone who has access to multiple cheap software-defined radios. Based on these technological developments, many crowdsourced sensor networks have sprung up, which collect air traffic control data in order to localize aircraft and visualize the airspace. Due to their unplanned and uncontrolled deployment and heterogeneous receiver technology traditional solutions to the Aircraft Localization Problem (ALP) can either not be applied or do not perform in a satisfactory manner.

In order to deal with this issue and to find novel approaches to the ALP itself, we have designed and executed a multi-stage open competition, conducted both offline and online. In this paper, we discuss the setup, experiences, and lessons learned from this Aircraft Localization Competition. We report from a diverse set of technical approaches, comprising 72 participating teams over three stages. The participants reached a localization accuracy of up to 25 meters in a setting with fully GPS-synchronized receivers and 78 meters in a largely unsynchronized receiver setting. These results constitute a significant improvement over the previous baseline used in the OpenSky research network.

We compare the results of the study, discuss the current state of the art, and highlight the areas that, based on our experience from organizing a competition, need to be improved for optimal adoption of the competitive approach for other scenarios.

\end{abstract}



\maketitle

\input{content}

\bibliographystyle{ACM-Reference-Format}
\bibliography{bibliography}



\end{document}

%% file: content.tex
\section{Introduction}

The opportunistic and independent localization of aircraft based on their wireless communication with other endpoints has been a longstanding problem and subject of much research. While modern aircraft generally have transponders, which they use to broadcast their position as obtained precisely via GPS or other Global Navigation Satellite Systems, there are a multitude of reasons to calculate or verify aircraft's positions independently on the ground. Among other applications, the localization of aircraft based on their communication signals can be used to track non-cooperative aircraft \cite{schafer2017opensky,Cycon17Strohmeier}, provide enhanced tracking information to legacy air traffic management systems \cite{strohmeier2014realities,schafer2018opensky}, improve the security of unencrypted ATC protocols \cite{strohmeier2020securing} or act as redundant backup and safety system in case of outages. In recent events, aircraft tracking has become an important factor in independently observing military movements in conflict zones, including the Ukraine war \cite{strohmeier2018utilizing}.

Originating from military settings, the capability to localize aircraft opportunistically has first expanded into the civil aviation domain, and is now available to any actor who controls multiple inexpensive software-defined radios (SDRs). The widespread proliferation of such SDRs has at the same time given rise to globally-oriented, crowdsourced flight information websites such as Flightradar24\footnote{\url{https://flightradar24.com}} and the OpenSky Network\footnote{\url{https://opensky-network.org}}. These organizations use the information gathered from many distributed SDR-based receivers of Air Traffic Control (ATC) data to display and share the tracks of aircraft around the world. The data of these networks is used for many critical applications, from air traffic management (ATM) to climate research and open source intelligence in times of conflict \cite{strohmeier2020research,buzzfeedosint}.

Consequently, the practical Aircraft Localization Problem (ALP) has expanded: from solving localized, controlled, homogeneous receiver environments to extremely heterogeneous, uncontrolled and global-scale crowdsourcing systems. However, the theoretical algorithms and solutions for use with the ALP have not been adapted to these developments.

Many classical solutions for the ALP have been proposed in the literature, with different theoretical underpinnings and typically focusing on the core multilateration algorithm. However, all suffer from two critical flaws for modern deployments. First, they were principally not designed for the crowdsourced case with its typical organic, non-controlled receiver growth and placement but instead rely on the ability to place receivers at will and in a proactive, near-optimal fashion. Second, until recently, there has been no scientific, standardized way to compare different methods of solving the different shapes and forms of the ALP. In learning from other areas of computer science such as operations and transportation research \cite{ropke2006adaptive} or network security \cite{van2018veremi}, general wireless \cite{dartmouth-campus-20090909} or internet data \cite{CAIDA}, this has recently been addressed with the LocaRDS dataset \cite{schafer2021locards}. This open dataset can be used for direct objective benchmarking of localization methods in aviation, which is our goal in the present paper.

Based on LocaRDS, we conducted an open competition in order to measure and improve the state of the art in ALP research and put it on a solid scientific grounding for the future. We aim to reduce the existing fragmentation of research on the ALP, where authors have to build their own test sets from real or simulated data in order to compare their novel methods. Naturally, as previously used data and metrics were generally not available and documentation is sparse, the reproducibility and comparability of results has been very limited until now. By using LocaRDS, we exploit the availability of open real-world crowdsourced flight data, which fulfills the requirements of different localization methods. In doing so, we hope that through the use of a comparable and standardized source, it becomes clear which are the best solutions to the ALP in different scenarios. 

Competitions using open and closed datasets have been a widely used method in several areas of computer science, particularly in machine learning and related areas. They have been shown to be a useful tool to engage the community, include stakeholders from outside academia and further a specific, often underdeveloped cause and also been applied successfully in indoor localization before, both in-person  \cite{lymberopoulos2015realistic} and online \cite{kaggle}.

\subsection*{Contributions}

\begin{itemize}

    \item We use LocaRDS \cite{schafer2021locards}, a reference dataset for effective scientific comparability in localization research based on crowdsourced real-world air traffic data, in order to derive an effective benchmark for the aircraft localization problem.
    
    \item We report the design and execution of a year-long public competition built to find novel and improved solutions to the ALP, in particular for the important crowdsourced setting.
        
    \item We analyze the impressive results of the participants and their technical design choices and distill lessons learned from our long-term efforts.
\end{itemize}

\section{Background}\label{sec:background}

\subsection{Aircraft Localization}

\begin{figure}
\includegraphics[width=0.95\columnwidth]{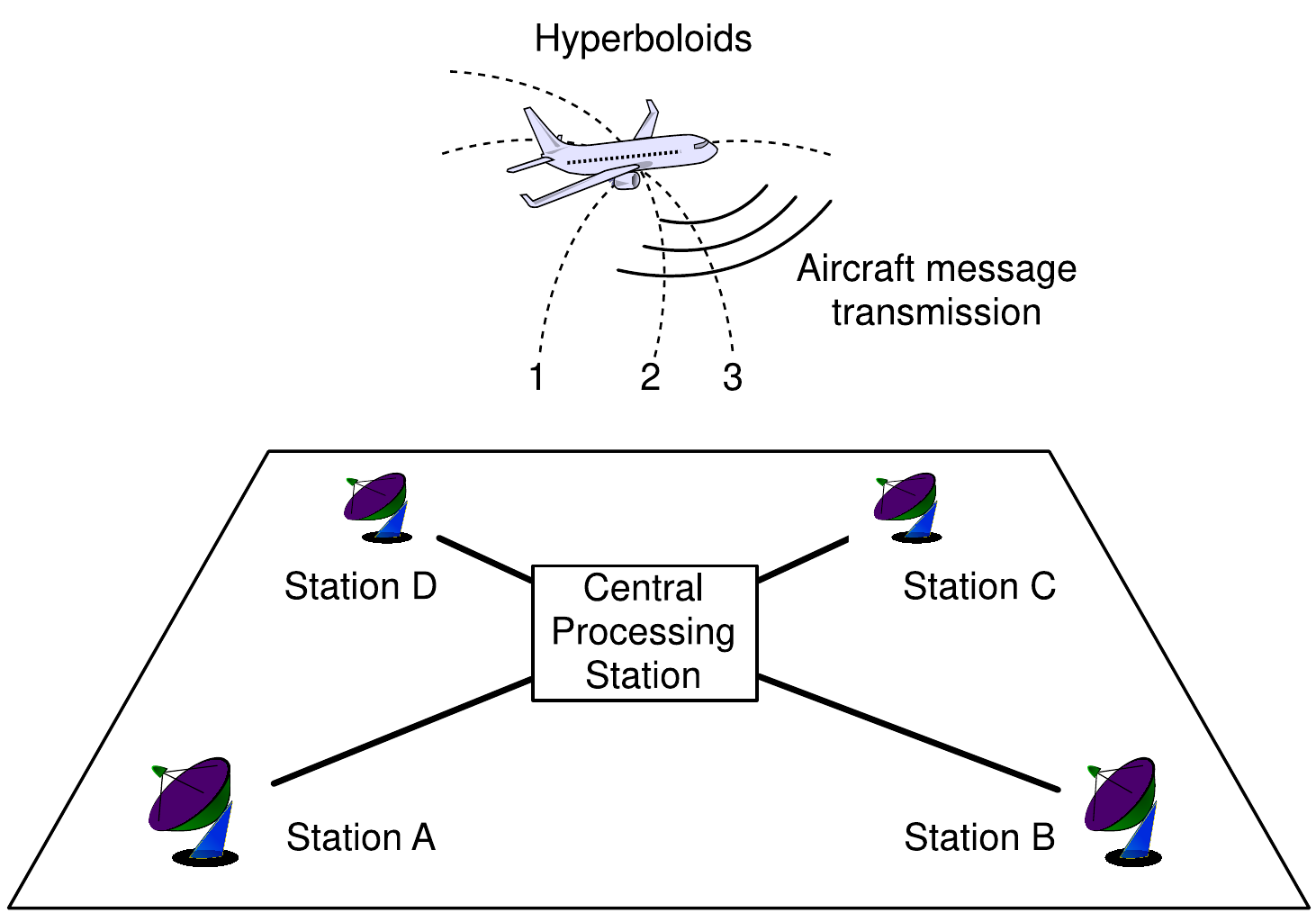}
\caption{Representation of the ALP.\label{fig:Representation-of-ALC} }
\end{figure}

First, we introduce the long-distance outdoor positioning problem of locating aircraft. We formulate this as the problem to find the 3D position $ \mathbf{p} = [x, y, z]^T$ of an aircraft based on the signal characteristics of the communication. Fig. \ref{fig:Representation-of-ALC} illustrates the process in abstract in the physical world.

This type of localization, or positioning, can in principle be conducted with any communications signal sent out by an aircraft, including analog signals such as VHF radio\footnote{Although obtaining accurate timestamps is complex in the analog case.}. Without loss of generality, LocaRDS uses the widely supported ADS-B technology, which is mandated in most developed airspaces from 2020 onwards and forms the heart of the next generation of air traffic control. ADS-B is readily collected by many web trackers including, as mentioned before, OpenSky or Flightradar24. Thus, it offers not only sufficient data but based on its popularity also many target users that would benefit from improved ALP solutions, in particular in a crowdsourced setting. 

\subsection{Existing Solutions for the ALP}

We discuss the two fundamental physical-layer characteristics used in the literature for solving the ALP, TDoA and RSS, in more detail. Both can be applied using LocaRDS and has been made available for use in the competition.

\subsubsection{Time Difference of Arrival}

The most popular approach to aircraft localization is to use the time differences of arrival concept, where $n > 1$ ground sensors receive and match the same signal sent by an aircraft. At reception, every receiver timestamps the signal. The ToA measurements are then joined and the differences of all arrival timestamps $t_1,...,t_n$ between all $n$ involved receivers calculated. This is done for example by subtracting the earliest timestamp $t_{min}$ or using a fixed receiver of the set as anchor. This data then forms the basis for the TDoA approach, which as a surveillance technique is best known as \textit{multilateration} (MLAT).

Calculating the position from this TDoA data is done traditionally through solving a geometric problem, i.e. finding the intersection of the resulting system of hyperboloids. Solutions have been proposed using iterative and closed-form direct algorithms,\footnote{A short classification of these methods can be found in \cite{SIVP}.} 
while the former require an initial estimation of the wanted position as input, the latter do not.

MLAT is a proven and well-understood concept used in civil and military surveillance. It serves as an operational method for ATC around airports and even smaller countries (e.g., Austria or Czech Republic). Academic works and aviation regulatory bodies have argued for MLAT being an ideal backup for primary radar systems, which are slowly being phased out due to cost, accuracy and reliability issues \cite{WAM}.

However, classic MLAT solutions suffer from drawbacks, most notably expensive hardware to enable highly accurate timestamps and tight synchronization. Both are a strict necessity for MLAT algorithms as they are highly sensitive to noise, in particular in uncontrolled receiver placements where the geometric characteristics are not optimized  \cite{strohmeier2016localization}.

\subsubsection{Received Signal Strength}

The alternative to TDoA-based localization is to directly determine the distance between a target and multiple reference locations. The target location is then estimated by finding the intersection of the resulting circles around the reference locations. 

The distances can be obtained by measuring a signal's time of flight or its strength. The former can be obtained through the round trip time (as in classical radar) or by relying on a tight clock synchronization between the target and the localizing infrastructure. Both methods are generally more limited in their applicability since they either require active communication or expensive synchronization of the targets. The RSS, on the other hand, offers a cheap alternative as it is measured by most available receivers.

By measuring the strength of an incoming signal and by knowing or estimating the transmit power of the aircraft, the distance between sender and receiver can be estimated based on their difference, i.e., the \emph{path loss}. One notable drawback of the RSS is, however, that its accuracy depends on many potentially unknown factors, the radio environment, and (analogous to TDoA) the measurement resolution \cite{Whitehouse07}.

Besides direct ranging measurements, RSS-based localization approaches often use indoor radio maps. This is intuitively more difficult to recreate with fast-moving aircraft spread out over long distances in highly-dynamic environments (due to weather, buildings and other influences). Building radio maps further requires a setup phase and separate infrastructure, which cannot be offered through a reference dataset. These are likely reasons why the RSS has been underutilized in our competition.

\section{Competition Design}\label{sec:design}

We discuss the setup of our competition(s), followed by the technical evaluation and the results.

\subsection{Goals}
We defined the design goals of the competition as follows:

\begin{itemize}

   \item \textbf{Batch localization: } Our competition mimicked an offline, batch localization problem, where a flight has fully finished and complete knowledge is available. Thus, exploiting the fact that aircraft move in predictable trajectories is explicitly allowed as is this the inclusion of ``future data''.  This is in contrast to live online multilateration, where only knowledge up to the present point can be included. We plan to examine this related problem in a future competition.

    \item \textbf{Target metric:} The target metric was exclusively the localization accuracy, here loosely defined as difference between the ground truth and the localization data.
    
    \item \textbf{Integration of LocaRDS: } We used an existing, comparable dataset that offered much training data for the participants, including both RSS and TDOA.
    
    \item \textbf{No computational requirements: } There were no requirements on execution time placed on the participants, in line with the batch localization goal. 

\end{itemize}

\begin{figure}[t]
    \centering
    \includegraphics[width=0.75\linewidth]{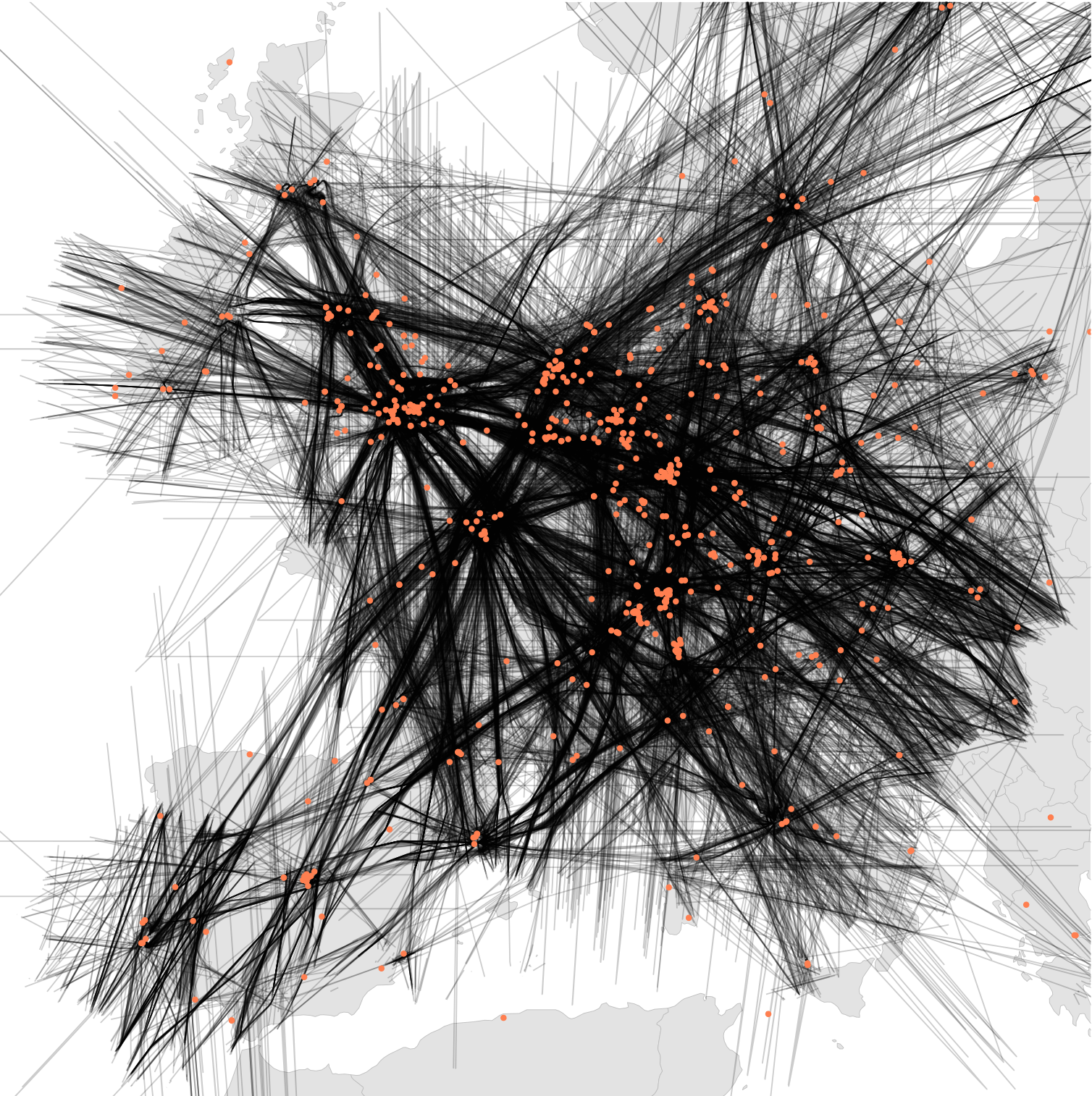}
    \caption{Illustration of the full LocaRDS dataset with 50,865,291 aircraft positions (black lines) and 323 sensor positions (orange dots). In addition to geographic information, the dataset contains time of arrival and signal strength measurements for each position reported by an aircraft.}
    \vspace{-5pt}
    \label{fig:locards_locations}
    
\end{figure}

\subsubsection*{Implicit Goal: Sensor Synchronization} \label{sec:sensor-sync}

Calibration and synchronization of the receiving sensors is effectively a pre-requirement for all practical localization methods, and it is of particularly crucial importance in the uncontrolled crowdsourced setting. It can thus be considered a separate, implicit subgoal of our competition. Many of the existing localization solutions require very tight time synchronization, in particular those based on TDoA measurements. This is costly even in controlled industrial deployments but impossible to achieve consistently with the variety of modern crowdsourced sensors used by enthusiasts to feed OpenSky and similar networks. While some algorithms may be more or less robust against noise in the TDoA data \cite{strohmeier2016localization}, the better the synchronization, the better the end results will be.

\begin{table}[]

     \caption{Synchronized competition datasets.}
     \label{tab:R1_stats}
     
\begin{tabular}{llll}

\hline
Offline / R1   & Positions & Flights & Size {[}MiB{]} \\ \hline
Training Set 1 & 2,074,194 & 2769    & 307.8          \\
Training Set 2 & 1,887,990 & 3076    & 277.7          \\
Training Set 3 & 2,002,847 & 2809    & 296.4          \\
Training Set 4 & 1,994,590 & 2585    & 300.0          \\
Training Set 5 & 1,951,877 & 2319    & 295.1          \\
Training Set 6 & 1,930,138 & 2347    & 296.9          \\
Training Set 7 & 1,869,587 & 2144    & 283.8          \\ \hline
Test Set       & 1,836,730 & 2888    & 272.9          \\ \hline
\end{tabular}
\end{table}

\subsection{Competition Datasets}

The offline competition and the first round of the online competition shared the same datasets (training and test) provided by the non-profit research network OpenSky with exclusively GPS-synchronized receivers. For the second round of competition, we used the open LocaRDS dataset \cite{schafer2021locards}, which included both synchronized and non-synchronized receivers from the crowdsourced OpenSky Network.\footnote{The competition ended before the full release of LocaRDS in March 2021.}

\begin{table}[]

     \caption{Non-synchronized competition datasets.}
     \label{tab:R2_stats}
\begin{tabular}{llll}
\hline
Round 2        & Positions & Flights & Size {[}GiB{]} \\ \hline
Training Set 1 & 6,535,444 & 2888    & 1.20           \\
Training Set 2 & 6,569,830 & 2818    & 1.20           \\
Training Set 3 & 6,348,679 & 2680    & 1.17           \\
Training Set 4 & 6,111,569 & 2932    & 1.11           \\
Training Set 5 & 6,309,260 & 2854    & 1.15           \\
Training Set 6 & 6,345,589 & 2812    & 1.16           \\
Training Set 7 & 6,187,378 & 2695    & 1.14           \\ \hline
Test Set       & 6,457,542 & 2929    & 1.18           \\ \hline
\end{tabular}
\end{table}

Fig. \ref{fig:locards_locations} illustrates the positions of the sensors and the measured aircraft trajectories. As can be seen, the focus of the dataset is on Europe, where the underlying data provider OpenSky has the best coverage with a sufficient number of sensors to conduct practical localization. Tables \ref{tab:R1_stats} and \ref{tab:R2_stats} provide the stats for the different test and training datasets. For the full detailed description of the features and an in-depth discussion of the design of LocaRDS, see \cite{schafer2021locards}.

\subsubsection{Sensor Dataset}

For all rounds the relevant sensor information was provided in a separate CSV file. These covered a subset of 514 (offline/R1) and 716 (R2) receivers, respectively, which were feeding aircraft data to the OpenSky Sensor Network in the relevant time period in 2018. The sensor data comprised their type, capabilities (GPS-synchronized or not) and the precise location as provided by OpenSky. It is worth noting here that the positions of the sensors are of varying accuracy. The sensor positions have only been entered by the user when the sensor was added to the network and there is no guarantee for correctness or accuracy. While some users report accurate positions for their antennas (e.g., measured with their smartphone), others just provide a rough estimate based on services like Google Maps. Some may even report wrong locations for privacy reasons.

\subsubsection{Training Datasets}

Each dataset (a CSV file) contained the data recorded by the OpenSky Network over a duration of 1h and has an (uncompressed) size of about 300MB (only synchronized receivers) or 1GB (all receivers). Each row represents the reception of one aircraft position report and contains the following information: a unique aircraft identifier, the Unix timestamp indicated when the message was received by OpenSky, unique identifiers of all sensors which received this signal, nanosecond timestamps from each of the sensors, signal strength measurements from each of the sensors, the position of the aircraft (latitude, longitude, height), the barometric altitude of the aircraft.

\subsubsection{Test Datasets}

The test or evaluation CSV was constructed in the same way as the training datasets. We then excluded the longitude, latitude and geometric altitude of an arbitrarily chosen 10\% of the flights, which the participants had to predict for the competition. This means the full position is present for all other aircraft in the dataset and can be used to synchronize the receiver clocks. Furthermore, the rough geometric height of the aircraft can be estimated based on the barometric altitude provided.

\subsection{Evaluation Metrics}
The metrics chosen for the scientific evaluation of the ALP should be as broadly applicably to the different scenarios and approaches as possible. In particular in case of a formal competition, they should further have as low a complexity as possible so the users can easily understand how they are calculated. Finally, they should be robust against cheating.

\subsubsection{Localization Accuracy}
The key metric for research in localization in general is the accuracy with which the position of the target is predicted. While the utility of aircraft localization depends on the context and the use case, more accuracy is strictly better.

The RMSE has been widely included as a standard metric to compare the predictive performance of different localization models (see \cite{lymberopoulos2015realistic}). 
However, as a basic metric, we chose the Truncated Root Mean Square Error (TRMSE) between the real aircraft position as reported by the ADS-B ground truth and the contestants' predictions for our main ranking metric. This makes the metrics more robust against a small number of outliers with large position errors.

\subsubsection{Dataset Coverage}

The second consideration concerns the coverage of the evaluation datasets, i.e., how many of the data points were chosen to be predicted. While ideally all samples would have a prediction, this is not practical for several reasons. For example, some methods may need initial samples to calibrate and also regularly re-calibrate. Furthermore, there is also value in correctly choosing to not predict bad or uncertain samples in order to minimize outliers and improve the average localization performance.  However, it is obvious that with equal localization accuracy higher coverage is strictly better.

Concretely, we first required a minimum sample coverage of 50\%, which should on average satisfy any non-tactical applications of the ALP, i.e. those where update rates of aircraft positional information of more than 1 second are allowed. However, other values can sensibly be chosen based on the application requirements and also depending on the sensor coverage in a given geographical region. 

\subsubsection{Further Considerations}
Due to the variation in the distribution of uncertainty and quality of measurements in OpenSky, it is clear that there can be trade-offs between coverage and accuracy, which we might want to capture. Besides requiring a minimum coverage, this trade off could also be quantified for a provided solution through applying a penalty directly towards the accuracy scoring. By assuming a fixed high localization error for any missing observation, the TRMSE is increased, incentivising the contestants to provide a higher number of observations. However, the effectiveness of the penalty is highly dependent of the quality of the provided solution --- if the penalty is set below the TRMSE, it will actually improve the quality score and thus set a false incentive to leave out observations. As we were not aware of the quality of these solutions, we dropped the application of such a penalty and do not report it.

A second consideration is centered around the runtimes of the provided solutions. While the speed of localization algorithms is not crucial in our batch localization scenario, it may still be insightful to analyze. Variations in training times for ML-based solutions may impact the choice of algorithms in situations where regular re-training is required. Similarly, lightweight algorithms for distributed resource-constraint edge computing are a relevant application for example for crowdsourced flight tracking networks. However, we decided to exclude the runtime from the initial scoring of the competition to enable participants to work on their own environments, making direct comparisons difficult.

\section{Competition Execution}\label{sec:execution}

We briefly discuss the first, offline, stage of the competition before providing more detail about the significantly more successful online stages.

\subsection{Offline Competition}

\subsubsection{Format}
We first decided to conduct this competition offline and in person, in conjunction with a leading academic conference on sensor networks. As the chosen venue regularly hosts on-site competitions of varying nature, this provided us with several advantages. First, a fixed framework with a pre-specified day and location. Second, embedding into a major conference would give additional awareness among a relevant academic community.

Participants from all backgrounds were free to join the contest, whether coming from academia, industry, government or out of private interest. Two months before the on-site meting, the competition attracted pre-registrations from 42 contestants with 33 different affiliations. 

As an additional incentive beyond the scientific challenge, there was significant prize money available. 

\subsubsection{Rules}
The competition required every solution considered for the awards to be open-sourced and their integrity subsequently verified by the organizers. Concretely, all source codes and additional datasets used to generate the results from the measurement data needed to be published under the GNU GPLv3 license. In addition, sufficient documentation needed to be provided to understand and reproduce the results.

Usage of any external datasets (e.g., weather data or tracking data from other sources) required explicit permission by the organizers one month prior to the on-site competition day and sharing with all other contestants. Contestants were only allowed to use their own original implementations. The simple re-use of existing code was explicitly disallowed. We encouraged individuals and teams of up to 5 persons from all backgrounds to register and participate. No affiliation with any of the organizers or their institutions was allowed. 

\subsubsection{Execution}

Three months before the day of the on-site competition, we provided the training datasets to all competitors in form of CSV files, which included the ground truth of all aircraft locations. These datasets could be used by the participating teams to train their models. For participating in the on-site competition, each team had to send at least one team member to the conference, where they received access to a non-labeled evaluation dataset. They could be supported by their team members remotely and a VoIP channel to the competition site was constantly available. Overall, 6 teams attended the evaluation day, with several thwarted last-minute due to visa and flight issues.

The teams then had 9 hours to find all locations of aircraft that are missing location information in the datasets. Every 3 hours, the teams had to submit their intermediate results (as a CSV file) to the organizers present. The organizers calculated an indicator of the accuracy of their solution and provide an intermediate ranking. After 9 hours, the teams submitted their final results and the final ranking is determined. It was possible to submit multiple times in each 3 hour slot.

\subsection{Online Competition}

\subsubsection{Format}

We provided the labeled training datasets and the test datasets at the start of the competition period. The task was again to predict all locations of the aircraft flights that were missing location information. Each team (or individual) submitted their results for both rounds during the competition periods as a CSV file of a defined number of rows, which was uploaded to the AICrowd website.\footnote{Concrete link withheld for anonymization purposes.} Afterwards, an indicator of the accuracy (the 90\% TRMSE) of their solution was immediately calculated and an intermediate ranking provided. When the competition time ended, the final ranking was determined using this leaderboard.

\subsubsection{Rules}
The rules with regards to open-sourcing, licensing, external data used and eligibility remained the same as in the offline competition. For full award eligibility, the quality of the solutions was required to be below 1000\,m TRSME. Between 1000\,m  and 5000\,m still half of the award money for a top 5 finish would be distributed, and none above 5000\,m. For each round, the full awards were set at 4000, 3000, 2000, 1000, 500 CHF/USD/EUR, respectively.

\subsubsection{Execution}

\paragraph{First Round: Synchronized}
The first round ran from June 15 - July 31, 2020. In this round, all provided data was from GPS-equipped sensors, which simplified things significantly as discussed in Section \ref{sec:sensor-sync}. This means that the competitors did not necessarily have to put any effort into sensor time synchronization in order to achieve practical results.

For this round, we instituted a minimum coverage requirement of 50\%. Thus, to be ranked at least 50\% of the missing aircraft positions must be provided. Based on the rankings at the deadline, the underlying code was shared with the organizers by the top 5 teams/participants. We verified and ran the code independently in order to ensure that the entries were in accordance with the competition rules. As there were no issues, the winners were confirmed and the awards distributed. 

Finally, we solicited feedback on the AICrowd forum that was set up for this competition. This resulted in several helpful comments by the participants in how to make the second round more engaging and remove some frustrations.

\paragraph{Second Round: Non-synchronized}
The second round ran originally from September 15 to October 31, 2020. We made three main changes to the first round:

\begin{itemize} 

\item We instituted a minimum coverage requirement of 70\% since the results were better than expected in R1 and did not suffer much when requiring higher coverage. 

\item We restricted submissions to 5 per day in order to reduce submission spamming.

\item We implemented a separate public and a hidden score in order to reduce over-fitting.  During the competition, feedback was provided on the scoreboard based on a fixed, arbitrarily chosen, 30\% of all aircraft trajectories that needed to be predicted. After the end of the competition, the scores on the full test dataset were calculated and shown. The winners of R2 were determined by this full ranking on the whole test dataset.

\end{itemize}

However, we found that participation was lower as was the eventual localization success, thus no awards were distributed. Building on our efforts, we again listened to the participants' feedback and refined the second round. Most notably, we released the open-source code of the winning entries of round 1 as well as additional description of the training data. We re-ran the second round from December 1, 2020 until January, 31, 2021 where it saw a good uptake comparable to round 1 with excellent localization success. After the same verification procedure conducted in the first round, the same amount of prizes were awarded and the new solutions also open-sourced on Github.

\section{Competition Results}\label{sec:results}

\subsection{Offline Competition}

Table \ref{tab:results-offline} shows the on-site results of the top contestants as well as our reference implementation.  The teams picked one, or a combination, of several fundamental solution techniques: machine learning, traditional multilateration, ranging using the received signal strength, statistical regression, and analyzing the distribution of the data and deducting the position based on the historical data gleaned from the training sets.

Coverage choices varied significantly between 50\% and 100\%. While the two (at least partly) ML-based solutions targeted the whole dataset, all other teams chose to stay close to the 50\% requirement, which is likely reflecting the difficulty of localizing a significant part of the real-world dataset measurements, even though they were from synchronized receivers. Yet, a combined ML / MLAT algorithm provided the best solution after 9 hours with 100\% coverage and a 90\%-truncated RMSE of 11,915.81 meters. Overall, we noted significant improvements throughout the evaluation day for all participants, making it likely that all approaches can be made more accurate.

In comparison, our reference implementation based on traditional multilateration shows that good aircraft localization results can be achieved with crowdsourced measurements based on cheap off-the-shelf hardware. It targeted measurements with at least 3 receivers, as is geometrically required for pure MLAT, and thus achieved a coverage of 45\% and a TRMSE of 682.38 meters.

\begin{table}[tb]
\caption{Localization results of the offsite competition (synchronized), compared to the multilateration reference implementation of OpenSky.}
\label{tab:results-offline}
\small

\begin{tabular}{@{}llrr@{}}
\toprule
\textbf{Rank}    & \textbf{Solution Type}         & \textbf{Coverage} & \textbf{TRMSE {[}m{]}} \\ \midrule
1         & Machine Learning / MLAT            & 1               & 11,915.81           \\ 
2         & RSS ranging           & 0.62               & 22,654.32     \\ 
3         & Distribution Analysis & 0.52             & 36,505.45           \\
4         & Distribution Analysis & 0.51              & 44,818.18           \\
5         & Regression            & 0.5              & 50,708.14           \\ \midrule

Reference & MLAT                  & 0.45               & 682.38                 \\ \bottomrule
\end{tabular}
\end{table}

Despite having three months preparation time with the training datasets, all on-site contestants were significantly less accurate than the traditional reference implementation based on multilateration used by OpenSky. This disappointing result shifted our approach towards an online competition to increase ease of participation and widen accessibility. 

\subsection{Online Competition}

As discussed to in the previous section, the offline competition results were not able to beat the existing reference implementations, partly by a wide margin. This changed significantly in the online competition, both in the first round (GPS-synchronized receivers only) as well as in the more difficult second round (all receivers, including non-synchronized ones). Table \ref{tab:results} summarizes the results of both rounds.

\begin{table}[]
\caption{Winning entries of the online competition.}
\footnotesize
\label{tab:results}

\begin{tabular}{@{}llllll@{}}
\toprule
\textbf{Rank} & \textbf{Team Type} & \textbf{Background} & \textbf{Coverage} & \multicolumn{2}{c}{\textbf{\begin{tabular}[c]{@{}c@{}}TRMSE [m] \\ (Public / Full)\end{tabular}}} \\ \midrule
\multicolumn{6}{l}{\textbf{Round 1 (synchronized, 50\% minimum coverage)}}                                                                                            \\
1             & Solo          & Academic         & 0.5               & 25.020                                       & -                                            \\
2             & Team          & Academic            & 0.502             & 25.817                                       & -                                            \\
3             & Team          & Independent         & 0.5               & 26.214                                       & -                                            \\
4             & Team          & Independent         & 0.502             & 33.544                                       & -                                            \\
5             & Solo          & Academic            & 0.5               & 59.467                                       & -                                            \\ \midrule
\multicolumn{6}{l}{\textbf{Round 2 (unsynchronized, 70\% minimum coverage)}}                                                                                          \\
1             & Solo          & Independent         & 0.7               & 78.14                                        & 81.89                                        \\
2             & Team          & Independent         & 0.7               & 90.13                                        & 98.37                                        \\
3             & Team          & Academic            & 0.7               & 141.07                                       & 154.57                                       \\
4             & Team          & Academic            & 0.723             & 157.32                                       & 171.66                                       \\
5             & Solo          & Academic            & 0.723             & 1497.99                                      & 2392.53                                      \\ \bottomrule
\end{tabular}
\end{table}

\begin{table*}[]
\caption{Methods used by the winning entries.}
\footnotesize
\label{tab:methods}

\begin{tabular}{@{}lll@{}}
\toprule
\textbf{Round 1} & \textbf{Pre-Processing }                                                                                                                                                           & \textbf{Post-Processing}                                                                                                                                                          \\ \midrule
\# 1       & Minimize offsets with training data.                                                                                                                                       & Fit spline to trajectory, identify good quality timings.                                                                                                                 \\
\# 2       & \begin{tabular}[c]{@{}l@{}}Correct offsets, exclude bad sensors, first guess using ML. \\ (Gradient Boosting Trees).\end{tabular}                                          & \begin{tabular}[c]{@{}l@{}}Filter for aircraft with sufficient sensor data (9 triplets), \\ smooth trajectories.\end{tabular}                                            \\
\# 3       & Calculate offsets.                                                                                                                                                        & Filter outliers, Hooke Jeves, linear interpolation, fit a/c track.                                                                                                  \\
\# 4       & Filter outliers.                                                                                                                                                          & \begin{tabular}[c]{@{}l@{}}Filter for direction and density with DBSCAN. \\ Localized extrapolation with Huber regression.\end{tabular}                                  \\
\# 5       & Identify good sensor combinations from training data.                                                                                                                     & Identify trajectories, filter outliers.                                                                                                                                  \\ \midrule
\textbf{Round 2} &                                                                                                                                                                           &                                                                                                                                                                          \\ \midrule
\# 1       & \begin{tabular}[c]{@{}l@{}}Estimation of effective signal wave velocity, including altitude dependency. \\ Iterative synchronization of good stations first.\end{tabular} & Huber Regression + Graph-based filter.                                                                                                                                   \\
\# 2       & Local adaptive sensor synchronization. Sensor outlier filtering.                                                                                                          & \begin{tabular}[c]{@{}l@{}}Filtering of predicted outliers. Track smoothing w\textbackslash low-pass filter. \\ Interpolation of points with \textless 4 measurements.\end{tabular} \\
\# 3       & Global sensor synchronization. Calculate measurements with 4+ sensors.                                                                                                    & Interpolate missing points.                                                                                                                                              \\
\# 4       & Barometric altitude estimation. Sensor synchronization + error minimization.                                                                                               & Basic filtering / trajectory smoothing.                                                                                                                                  \\
\# 5       & Models clock drifts of sensors from training data.                                                                                                                        & Interpolates between predictions.                                                                                                                                       
\end{tabular}
\end{table*}

\subsubsection{First Round}

The results in the first round significantly beat our expectations in terms of accuracy. The three top results clustered within around one meter of 25\,m TRMSE, with the 4th and 5th place still below 60\,m. It can be noted that all teams perfected the coverage requirement of 50\%, selecting the most reliable data points only for grading and ranking. This is an intended feature that is also relevant in real-world localization systems.

All solutions were provided in Python/Cython and used fundamentally a variation of a classical MLAT approach. They differed, however, in their pre- and post-processing (see Table \ref{tab:methods}). Common themes included the identification of the most reliable receivers and the most accurate measurements and tracks. These were then filtered and smoothed using a wide variety of methods from DBSCAN to Hooke Jeeves.

We further observed that all teams are exactly at, or very close to the minimum coverage requirement. This illustrates the effort put into the optimal data selection.  `Sniping' was a practical issue we observed. As the deadline drew closer more teams entered the competition and significantly more results were entered for grading, leading to a frantic final day and deadline experience. In terms of team types, we had a mix of participants with academic affiliation and independent competitors, most were teams of two or three but the top contestant was a single individual.

\subsubsection{Second Round}

The second round also exceeded our expectations significantly. As only about 15\% of the available sensors were GPS-synchronized, this posed a much harder challenge. Which the participants solved with two excellent results below 100\,m and two below 200\,m. R2 was again won by a solo participant, this time without academic affiliation. 

Participants again approached the 70\% coverage requirement. The difference between public and hidden score was between 4.8\% and 9.6\% for the top 4. This means no significant overfitting and it did not affect the final ranking order.

Methodologically, the focus was on the accurate sensor synchronization, which the participants attempted to do either locally or globally and with different choices of good sensors. Interestingly, the winning solution, which will be discussed in detail in Section \ref{sec:winning}, incorporated open-sourced ideas from R1.

\section{Technical Evaluation}\label{sec:technical-evaluation}
In this section, we discuss the technical outcomes of the competition more deeply. To do this, we first recap the problem of multilateration and sensor synchronization before describing the reference solution that was used by the OpenSky Network as a baseline. We then analyze and contrast the methods of the top 5 participants before describing the winning solution in closer detail. As the RSS was not included by the top teams, we exclude it from here on.

\subsection{MLAT: Problem Statement and Solution}

A localization problem consists of relating the target position $\boldsymbol{\theta}=(x,y,z)^T$ with a set of observations $\mathbf{m}=(m_1,...m_M)$:
\begin{align}
    \mathbf{m}=\mathbf{f}(\boldsymbol{\theta})+\mathbf{n}
    \label{eq1}
\end{align}
where $\mathbf{n}$ is the measurement noise of the observations.
If the system can be solved, the target position can be estimated.

In case of TDOA measurements, obtained by subtracting the TOAs of the $ith$ receiving station and a reference one (for example the first station), the $\mathbf{f}$ represents $M$ hyperboloids and it can be expressed as follows: 
\begin{align}
    m_i=TOA_i-TOA_1=\frac{\|\boldsymbol{\theta-\vartheta_i}\|-\|\boldsymbol{\theta-\vartheta_1}\|}{c} +n_{i,1} 
    \label{eq2}
\end{align}

where $c$ is the velocity of light, $\mathbf{\vartheta_1}=(x_i,y_i,z_i)^T$ represent the sensor position, and $n_{i,1}$ represents the difference between the realization of the noise of the sensor $n$ and the sensor $1$.
In general, it is possible to characterize this noise with its covariance  matrix $\mathbf{N}$.

It follows that the localization performance (generally referred as localization accuracy) depends on three elements: 
\begin{enumerate}
    \item the measurement accuracy (i.e. matrix $\mathbf{N}$), 
    \item the spatial distribution of the stations (that gives the  $\mathbf{f}$ function),
    \item  the algorithm used to solve Eq. (\ref{eq1}) for $\boldsymbol{\theta}$. 
\end{enumerate}

For the first two elements (measurement noise and sensor distribution) nothing can be improved with a crowdsourced network: it organically grows without any particular optimization of sensor positions, and exploits receiving hardware of a given, highly variant, performance. For these reasons, competitors can only pre-process the data to statistically characterize the measurement noise of the sensor and $\mathbf{N}$, to discard outliers, or to select the optimum subset of sensors.

Concerning the algorithm used to solve the inverse problem, typically, the system  in Eq.\,(\ref{eq2}) can be solved with a number of different and well-known approaches:

\begin{itemize}
    \item It can be treated as a regression problem, which can be linearized and solved using classical Least Squares (LS). This is the most common approach in literature and practice but requires an initial guess of the position. It also does not take into account a priori information about the mobile kinematics (iterative solution).
    
    \item It can be treated as an statistical estimation problem and any classical estimator (e.g., the Maximum Likelihood Linear Estimator or MLLE) can be applied knowing the distribution of $n_{i,1}$. If the error model is well suited, this approach usually gives non-biased solution 
    close to the Cramer-Rao Lower Bound. Under some simplification this approach leads to the LS solution. In any case, an initial guess of the solution is needed (iterative solution).
    
    \item Some numerical methods set a new mathematical function that relates the unknown target position, the measurements, and a new parameter derived from the target position (e.g., the target range). The resulting models are usually linear in one unknown; given the other ones and, assuming certain numerical approximations between the target position and the derived parameter, is possible to compute the solution.  These methods do not require any initial guess to work and are commonly referred as closed form algorithms. On the other hand, they usually introduce quadratic noise terms in the inverse problem, and usually the solutions are biased. Examples of this approach are shown in \cite{smithabel,Schau1987,chanho}.  
    
    \item Geometrical methods that algebraically manipulate the hyperbolic equations until they directly set an inverse problem relating the target position with the measurements. These models usually require more measurements, introducing also quadratic and cubic noise. As the numerical methods, these do not require any guess for the solution (closed form) but typically are biased. Examples are given in \cite{Schmidt1972,Geyer1998,Leonardi2009,Leonardi2008}.
    
    \item Recently, machine learning approaches have been proposed, which do not rely on inverting the system in Eq.\,(\ref{eq1}). A data set with all possible measurements, computed on a grid of points is generated and than the best match between the incoming measurement vector and the dataset is found by the use of a KNN algorithm \cite{Strohmeier15}. This approach can be classified as a fingerprinting classification and similar approaches were already tested in other application field such as indoor localization.
\end{itemize}

It must be noted that all approaches (except for the last one) require an inversion of a system of non-linear equations that must be numerically solved and, sometimes, it can be strongly ill-conditioned and not always numerically stable,  providing unacceptable accuracy. 

Fortunately, some numerical algorithms that regularize the inverse problem exist, such as the Tikhonov regularization, Singular Value Decomposition, and Total Least Squares (TLS)-based regularization, which have already been applied to the MLAT problem \cite{mantilla1,MANTILLA2013}. 

Finally, the methods that require an initial guess can be very sensitive to its choice. The solution can result in large errors if it is far away from the true target position. This problem is even more important in the ALP due to the fact that the considered scenario is global, contrary to typical local MLAT deployments. Thus, for any iterative method, the strategy to select the initial guess must be defined.

\subsection{Pre-processing and Synchronization}

Equation \ref{eq1} and \ref{eq2} assume that any station is synchronized with the reference station. Usually, time synchronization of the MLAT systems is achieved in two different ways: either with GPS/GNSS time distribution (integrating a GPS receiver on each sensor) or via a reference transponder in known position that transmits RF  messages to all sensors. Knowing the reference transponder position, it is possible to compute all the sensor clocks and synchronize the network. 

Both methods are impractical in crowdsourced networks. In this scenario, there are only a few sensors using costly GPS synchronization, while the largest part feeds data without any synchronization mechanism.

A common, sub-optimal, solution to overcome this fragmentation is the use of opportunity traffic --- airplanes transmit their position encoded in the ADS-B messages. This means that if the airplane is in view from more sensors, the time  biases between the stations can be easily estimated inverting the equation:
\begin{align}
    m_i=\frac{\|\boldsymbol{\theta-\vartheta_i}\|-\|\boldsymbol{\theta-\vartheta_1}\|}{c} +b_{i,1}-b_1+n_{i,1} 
    \label{eq3}
\end{align}
where $b_{n,1}$ represents the bias of the station $i$ w.r.t. the reference station $1$, and all position in the equations are known.
This method has some limitations: it is difficult to achieve the synchronization of the complete network and the synchronization performance depends on the sensor measurement noise, the sensor position error, the aircraft ADS-B position accuracy, and the system geometry.

Moreover, the estimation of the clock offset at one moment usually is not sufficient, due to clock drift over time.

Clock drift comes from two main components of error: systematic fluctuations and random fluctuations \cite{galleani_tavella}.
Systematic deviation over time can be written as follows (approximating to the second order term) \cite{galleani_tavella}:
\begin{equation}
b_{sys}(t)= b(0)+f(0)t+0.5Dt^2
\label{eqclock}
\end{equation}
where $b(0)$ is the initial time offset of the clock, $f(0)$ is the initial frequency offset of the clock and $D$ is the frequency drift of the oscillator (it represents the systematic change of frequency due to a combination of internal factors such as aging or production tolerances).

There are several known solutions to this problem:

\begin{itemize}
    \item Sequential estimation using a priori assumptions about the clock dynamic model and its noise characteristics (for example using Kalman filtering as done in \cite{kalman} or a simple alpha beta filter \cite{brookner}. The clock model and statistical properties or the errors are required.
    \item Regression to compute the parameters in Eq.\, (\ref{eqclock}). Only the regression formulation is required.
    \item Fitting/smoothing the sequence of measurements. No a priori information about the clock is required.
\end{itemize}

The first two methods use past data to extrapolate the future and can be used also in real-time applications. The third method is suitable only for offline batch localization.

Further pre-processing activities mentioned in the previous sections: statistical derivation from the data to be used for the target position by estimators, sensor characterization, outlier detection, data cleaning, sensor subset selection etc.

\subsection{Post-Processing and Aircraft Tracking}

Having computed the aircraft positions for each received message individually, it is possible to significantly improve the accuracy and completeness of the results by applying post-processing techniques. Considering that generally the trajectories of airplanes are smooth, or in accordance with dynamic constrains, it is possible to again use estimators to improve the quality of the final localization output. In particular, we can detect outliers, smooth the trajectory, interpolate the trajectory, or produce an initial guess for the following measurements. Known methods include sequential estimation by using a  priori statistical information of the aircraft and filtering as proposed by Kalman, or fitting and smoothing methods such as regression that can be used without a specific knowledge of the aircraft dynamic.

\subsection{Reference Solution}

The reference solution use the simple MLAT Least Square formulation, using all the pairs of sensor available for each ADS-B message. 
An initial guess for each solution is needed: it is computed simply by the weighted average of the receiving sensors coordinates or (if available) from the last computed position for the aircraft.

The pre-processing manly consists in time synchronization.
ADS-B opportunity traffic is used to synchronize pairs of sensors receiving the same ADS-B message, without looking for a total synchronization of all the stations to a common reference frame. 
This solution was chosen for simplicity and robustness: it is tolerant  to  stations  switching on and off in the time.
The synchronization method continually (and sequentially) estimate the clock offset and drift by the use on an alpha/beta filter (a simplified version of a Kalman filter) and, for each pair of stations is also able to predict the values for the next measurements. The prediction is also used to drop outliers that shows big steps in the estimated bias.

Post-processing is not applied: in this way the raw results and the real power of multilateration algorithm can be appreciated. 
The Reference solution is suitable both for real time and batch application because use the sequential approach for time synchronization, more over is able to re-start time synchronization also in case of lack of data for short or long period for any couple of paired stations.

\subsection{Competition Solutions}

One of the goals of the competition was to find new strategies (or new combination of strategies) to solve the MLAT problem in a demanding unsynchronized real environment.

The winning solution (R2), discussed in detail below, uses the training data to estimate the fixed measurement offset in the sensors, and to estimate the speed of the signal in the troposphere (instead of using the classical approximation of the speed of light). In post-processing, the trajectories were smoothed with a spline algorithm. This final step is the major contribution to improve accuracy and availability of the solutions. It is noted that the proposed method is not suitable for any real-time application.

Another interesting approach is a different way of obtaining local (pairs) synchronization used by the second classified teams: they use (instead of the classical sequential algorithms) an heavier linear regression of neighbors points to compute offset and drift terms (it doesn't care about the system model but, anyway, it  can be also used in real time application).
Concerning the MLAT solution, a closed form approach (generally lighter and simpler than the LS or MLE approach) was used and all the refinement is than demanded to the post-processing.

An important different approach was to first use linear regression for time synchronization, then solving the MLAT problem with Least Squares for subsets of four stations. This involved selecting the best subset to carry out the solution (this is a sort of sanity check of the measurements similar to integrity monitoring process in GNSS field). Additionally, it computes  the aircraft altitude using Light Gradient Boosting Machines because the MLAT problem is usually ill-conditioned in the vertical dimension and the inversion usually produce large errors in altitude.

Finally, while the obtained results are an order of magnitude worse than the top four, a radically different ML approach is proposed by the 5th-placed team. After global synchronization using linear regression, a grid of possible positions is defined and a cost function on this grid is minimized. This circumvents inverting the problem (similar to \cite{Strohmeier15}). 

\subsection{Winning Solution}\label{sec:winning}

We describe in detail the winning solution of the most complex R2 setup with unsynchronized sensors. It is based on solving classical MLAT equations both for aircraft track reconstruction and sensor synchronization using the test and the training datasets correspondingly. In comparison with Eq.\,(\ref{eq2}), two simple modifications were made:
\begin{itemize}
    \item The 1-norm was used due to presence of a large number of outliers in the time measurements and the fact that the measurement noise is not Gaussian.
    \item The velocity of light was replaced with the effective signal velocity taking into account refractive index of the atmosphere.
\end{itemize}

Thus, the following system of non-linear MLAT equations is considered:
\begin{equation}
    m_i = \frac{|\boldsymbol{\theta - \vartheta_i}| - |\boldsymbol{\theta - \vartheta_1}|}{\hat{v}} + b_{i,1} - b_1 + n_{i,1}
    \label{winning_mlat}
\end{equation}
where $\hat{v}$ is the effective signal velocity. These equations were solved using the Broyden–Fletcher–Goldfarb–Shanno algorithm (BFGS), which is a popular quasi-Newton optimization method for parameter estimation.

\subsubsection{Signal Velocity Model}

The radio wave velocity depends on the refractive index of the atmosphere and, therefore, on altitude:
\begin{equation}
    v(h)=\frac{c}{n(h)}=\frac{c}{1+A_0\cdot e^{-B\cdot h}}
    \label{signal_velocity}
\end{equation}
where $c$ is the velocity of light, $h$ is altitude, $n(h)$ is the refractive index, and $A_0$ and $B$ are some constants  \cite{Purvinskis2003}. The effective signal velocity (which is the average signal velocity along a signal path) can be derived from Eq.\,(\ref{signal_velocity}) as follows:
\begin{equation}
    \hat{v}\triangleq \frac{L}{\int{dt(h)}}=\frac{c}{1+\frac{A_0}{B\cdot\left(h_2-h_1\right)}\left(e^{-B\cdot h_1}-e^{-B\cdot h_2}\right)}, h_2 > h_1
    \label{effective_velocity}
\end{equation}
where $L$ is the signal path, $h_1$ and $h_2$ are altitudes of a receiver and a transmitter.

Though the radio wave velocity in the atmosphere is close to the velocity of light, the difference cannot be considered negligible in the real-world setting of the competition. For example, for $h_1$ = 0 km (Earth surface) and an aircraft altitude of $h_2$ = 10 km, the ratio $\hat{v}/c$ is equal to 0.9997 and the corresponding time-of-flight difference is 0.5$\sigma$ when an aircraft is at station's zenith and 5$\sigma$ when the distance between them is 100 km ($\sigma$ = 20\,ns is the standard deviation of the measurement noise).

\subsubsection{Pre-processing}

An initial (core) set of 36 GPS-syn\-chronized stations was selected. Their time measurements did not experience clock drift, but only fixed time offsets. Thus, the goal of the pre-processing step was to update stations' locations $\vartheta_i$ and determine stations' time offsets $b_{i,1}=const$ and two parameters $A_0$ and $B$ for the signal velocity model. For that purpose, Equations (\ref{winning_mlat}) with unknown variables $\vartheta_i$, $b_{i,1}$, $A_0$ and $B$ were solved on a subset of the training dataset by minimizing the sum of residuals. The initial guess for the BFGS algorithm was chosen as:
\begin{itemize}
    \item $\vartheta_i$: as reported stations' locations;
    \item $b_{i,1}$: as median values of time offsets;
    \item $A_0$ and $B$: both $10^{-3}$
\end{itemize}

\subsubsection{Synchronization}

All stations were synchronized with the network or excluded (in case of large fitting errors). Timestamps of a synchronized station can be expressed via aircraft time and vice versa which allows the propagation of absolute time values across the network:
\begin{equation}
    t^{meas} = t^{aircraft} + \frac{L}{\hat{v}} + b(t^{meas}), b(t) = 0
    \label{eq:aircraft_time}
\end{equation}
where $b(t)$ is the clock drift.

The initial set of 36 stations was already synchronized in the pre-processing step. All the other stations were considered to have clock drift which was modeled as a combination of a systematic fluctuation (Eq.\,(\ref{eqclock}) with linear order term only) and a random walk (fitted by a cubic spline):
\begin{equation}
    b(t) = b(0) + f(0)t + rw(t)
    \label{eq:winning_clock_drift}
\end{equation}
where $rw$ is a random walk.

After inserting Eq.\,(\ref{eq:winning_clock_drift}) into Eq.\,(\ref{eq:aircraft_time}), the following formula was derived to correct (synchronize) the time measurements of a station with clock drift:
\begin{equation}
\label{eq:correct_time}
    t^{aircraft} + \frac{L}{\hat{v}} = \frac{t^{meas}-b(0)-rw\left(\frac{t^{meas}-b(0)}{f(0)+1}\right)}{f(0)+1}
\end{equation}
Once $t^{aircraft}$ timestamps are calculated using synchronized stations, the clock drift parameters $b(0)$, $f(0)$ and the spline $rw(t)$ can be determined by minimizing the sum of residuals in Eq.\,(\ref{eq:aircraft_time}). After applying the right part of Eq.\,(\ref{eq:correct_time}) to station timestamps, it becomes synchronized with the network.

\subsubsection{MLAT Solution}

The target aircraft locations in the test dataset were calculated independently one-by-one by solving MLAT equations (\ref{winning_mlat}) for data points where three or more time measurements were available. The initial guess for each location was computed as barycentric position for a given set of stations. The aircraft track was considered as a directed graph with nodes in the estimated locations. The predicted locations were filtered by computing the maximum spanning tree of the graph where the aircraft velocity between any two nodes did not exceed 300 m/s.

\subsubsection{Post-processing}

Many target locations having less than three measurements or those filtered out in the previous step can still be reconstructed. Assuming smoothness of aircraft trajectories, latitude and longitude values of predicted locations were fitted by second order polynomials of aircraft time in local neighborhood. After that all locations (including those used for fitting) from the same neighborhood were reconstructed using the fitted polynomials.    
Additional splines of order 5 as functions of aircraft time were fitted to latitude and longitude values of predicted locations before reconstruction. These splines were used to estimate prediction accuracy of reconstructed points. The ones with the highest estimated errors were removed from the solution.

\begin{figure}
\includegraphics[width=0.95\columnwidth]{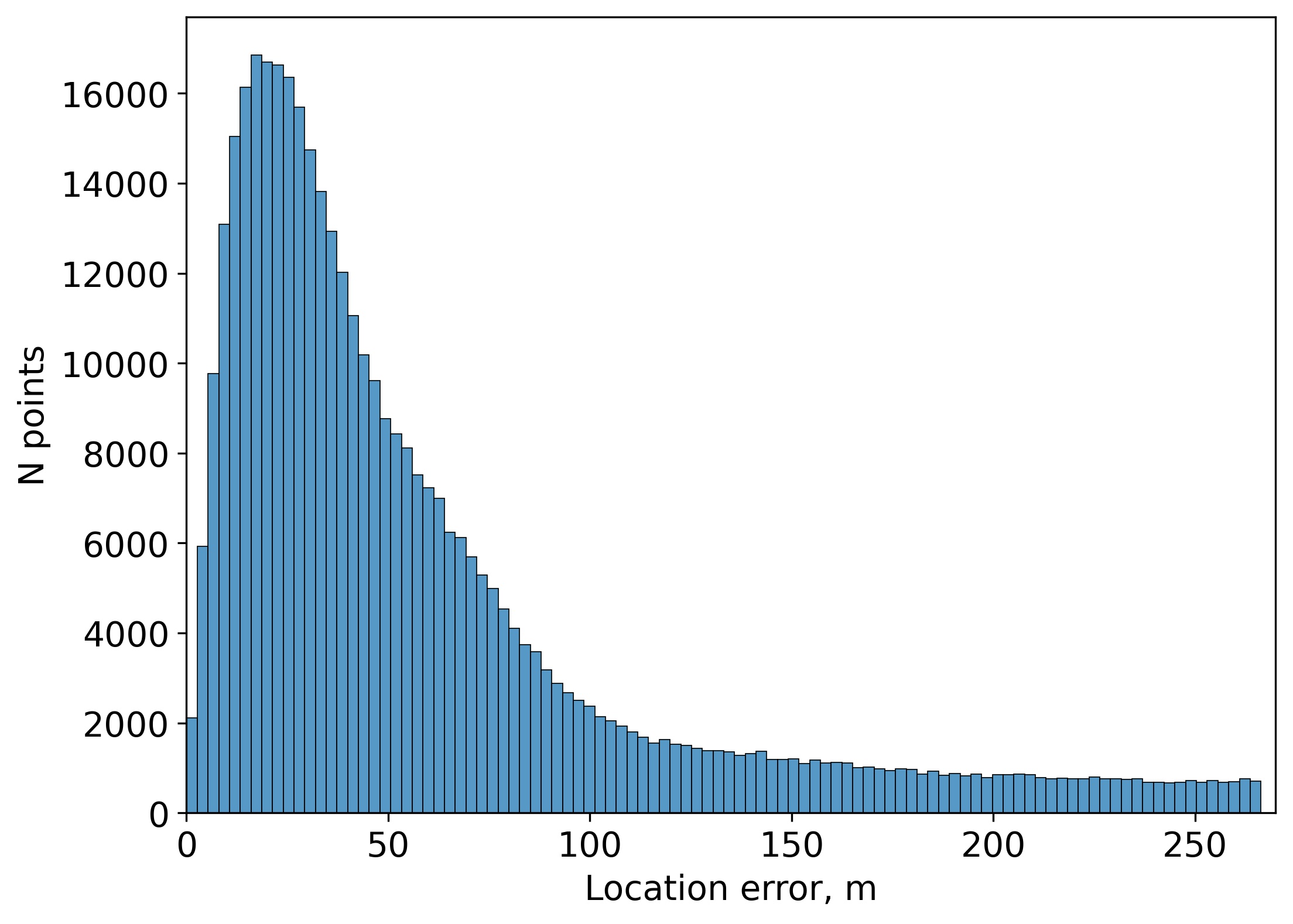}
\caption{Distribution of aircraft location errors obtained for the full LocaRDS dataset in the winning solution. \label{fig:Location-error-distribution} }
\end{figure}

Finally, it was found that filling short trajectory gaps (60\,s or less) with spline predictions improved total prediction accuracy. The winning solution showed 78.14m and 81.89m TRMSE on the public and the full LocaRDS datasets correspondingly, both with 70$\%$ of coverage. Distribution of location errors for the full dataset is presented at Fig. \ref{fig:Location-error-distribution}.

\section{Lessons Learned}\label{sec:lessons}

There are several insights from our process of designing and executing a large-scale competition on aircraft localization:\\

\begin{enumerate}
    \item \textbf{Offline vs. online competition:} There are several practical drawbacks of running on-site data science competitions, which from our experience outweigh the advantages even in the pre-pandemic world. These range from the costs and environmental footprint of experts traveling from around the world to issues with visas and last-minute weather-related cancellations. While the appeal of a live competition is tempting and face-to-face exchanges facilitate creativity and foster competition, these features do not compensate for the sheer efficiency of online contests. Our online competitions saw both a significant increase in contestants and improved results in the orders of magnitude.
    
    \item \textbf{Obstacles for industry participation:} We overestimated the interest of industry entities involved with multilateration or other localization approaches in an open competition. While in theory, competitions can be attractive to companies as they can show off the quality of their work, it is also a risk factor in the marketplace in case their solution performs worse. Another significant factor that prevents industry is the protection of intellectual property surrounding the solution or the implementation, even when the source code does not need to be opened.
     
     \item \textbf{Value of open-sourced localization code:} Our goal of advancing practical aircraft localization and making it accessible beyond the proprietary industry systems, where it is currently prevalent, included the open-sourcing of the code from the very beginning. Our experiences with the extension of round 2 illustrate the power of this open and collaborative approach; all top results improved significantly, often by learning from the published approaches (code and documentation) of round 1. In the mean time, several new projects have popped up on Github, illustrating their own methods and building on the code published after the conclusion of R2.
     
    \item \textbf{Improving existing algorithms:} Analyzing the different solutions paved the way for new localization research, merging insights from different fields of research, such as machine learning and geometrical solutions. Moreover, it was clearly shown that a big improvement of the ALP solution can be obtained applying well tailored post-processing and data smoothing.
    
    \item \textbf{Ongoing evaluation} The widely varying approaches both in the literature and our competition show that there is still much room for improvement in the theoretical development and practical implementation of solutions for the ALP. The results show that the contestants have significantly improved on the OpenSky reference implementation. We expect further improvements through open research with the released data, code, and scientific reports to a wider audience.
\end{enumerate}

\section{Conclusion}\label{sec:conclusion}

Crowdsourced air traffic trajectories are used in many areas of science and commerce and depend on the quality and accuracy of the observed data. In this paper, we have presented the design and execution of a multi-stage open competition on solving the aircraft localization problem in this context. The 72 participating teams reached a highly practical localization accuracy of up to 25 meters in a fully GPS-synchronized setting and 78 meters in a largely unsynchronized setting with the cheapest possible receiver hardware (50 USD and less). By analyzing comparing online and offline competitions many novel lessons were learned for future scientific challenges, including real-time localization.

\section*{Code and Data Availability}

The complete code of the aircraft localization competition, including the winning entries, has been made available at \url{https://github.com/openskynetwork/aircraft-localization}. The training and test data has been made permanently available on Zenodo: \url{https://zenodo.org/record/4739276}. The competition websites are available at AICrowd (\url{https://www.aicrowd.com/challenges/cyd-campus-aircraft-localization-competition/}) and the OpenSky Network Association (\url{https://competition.opensky-network.org/}).